\documentclass{article}
\usepackage{amssymb}
\usepackage{amsmath}
\usepackage{amsthm}

\newtheorem{theorem}{Theorem}[section]

\numberwithin{equation}{section}

\begin{document}

\title{Mathematical aspects of the cold plasma model}
\author{Thomas H. Otway\thanks{%
email: otway@yu.edu} \\
%EndAName
\\
\textit{Department of Mathematics, Yeshiva University}\\
\textit{\ \ New York, NY 10033, USA}}
\date{}
\maketitle

\begin{abstract}
A simple model for electromagnetic wave propagation through
zero-temperature plasma is analyzed. Many of the complexities of
the plasma state are present even under these idealized
conditions, and a number of mathematical difficulties emerge. In
particular, boundary value problems formulated on the basis of
conventional electromagnetic theory turn out to be ill-posed in
this context. However, conditions may be prescribed under which
solutions to the Dirichlet problem exist in an appropriately weak
sense. In addition to its physical interest, analysis of the cold
plasma model illuminates generic difficulties in formulating and
solving boundary value problems for mixed elliptic-hyperbolic
partial differential equations.
\end{abstract}

\section{Introduction}\label{otsec1.1}

Among the many equations of mathematical physics which change from
elliptic to hyperbolic type along a smooth curve, only the
equations for transonic flow have received sustained attention. In
this brief review we consider elliptic-hyperbolic equations
originating in a simple model for the propagation of
electromagnetic waves through zero-temperature plasma. Solutions
to such equations are likely to have significantly weaker
regularity than solutions to the linearized equations of transonic
flow. Recognizing the interdisciplinary nature of the topic, we
assume a familiarity with physics but not necessarily plasma
physics, and analysis but not necessarily elliptic-hyperbolic
equations. However, the physics is confined to a review of
fundamental results in Sec.\ \ref{otsec1.2}, whereas the
mathematical results of Sec.\ \ref{otsec1.3} are somewhat more
technical. There we consider the extent to which problems
formulated primarily for linearized equations of gas dynamics
possess analogies for equations arising from a different physical
problem. Continuing such investigations in various physical and
geometric contexts (\emph{c.f.} \cite{otO3}), one may
hope to obtain eventually a natural theory for linear
elliptic-hyperbolic partial differential equations.

\subsection{Physical background}\label{otsec1.1.1}

The plasma state is characterized by the dominance of long-range,
nonlinear effects. For matter in such a state, it is particularly
difficult to obtain mathematical problems which can be stated with
a satisfactory degree of rigor, and for which solutions can be
shown to exist. Without a proof of the existence and uniqueness of
solutions --- which, in particular, specifies the function spaces
in which solutions lie --- it is hard to place appropriate
boundary conditions on numerical experiments and to gauge the
reliability of the results obtained.

If one hopes to obtain a tractable mathematical problem, it is
usually necessary to impose harsh assumptions on both the plasma
and the applied field. Perhaps the harshest of these fixes the
temperature of the plasma to be zero. This permits one to neglect
altogether the fluid properties of the medium, which is then
treated as a linear dielectric. Somewhat surprisingly, the
assumption of zero plasma temperature is a useful first
approximation to the products of tokamaks: low-density plasmas
which are remarkably free of expected high-temperature phenomena
such as collisions and wall effects. See the remarks in the
introduction to  \cite{otSt} and the more detailed
discussions in  \cite{otW1}. More generally, the cold
plasma model approximates the effects of small-amplitude
electromagnetic waves, propagating with phase velocities which are
sufficiently large in comparison to the thermal velocity of the
particles.

We note that the term \emph{cold plasma} is highly ambiguous.
Although we take this to imply zero temperature, in the
astrophysics literature interstellar plasmas on the order of
$10^4$ $K$ to $10^5$ $K$ are typically referred to as ``cold"
(see, \emph{e.g.},  \cite{otCG}). Very recently,
``ultracold" neutral plasmas, having electron temperatures in the
range from 1 $K$ to $10^3$ $K$ and ion temperatures ranging from
$10^{-3}$ $K$ to 1 $K,$ have been created experimentally. The cold
plasma model explored in this paper is apparently too simple to
yield quantitative insight into those plasmas. In particular, the
fluid dynamics aspects of experimental ultracold plasmas appear to
be non-negligible (\emph{c.f.} Sec. 3 of  \cite{otKPPR}).

The other physical hypotheses imposed in this review are also
quite restrictive: Although the plasma is not assumed to be
homogeneous, the inhomogeneity is taken to be two-dimensional, so
the governing equations for the model are also essentially
two-dimensional. Moreover, the applied magnetic field is assumed
in Sec.\ \ref{otsec1.2.4} to be longitudinal and the resulting
wave motion confined to electrostatic oscillations. In Sec.\
\ref{otsec1.2.5} we consider electromagnetic waves, but we find
(after reviewing a detailed analysis by Weitzner \cite{otW1}) that
elliptic-hyperbolic equations arising in the electrostatic case
retain their validity as a qualitative model for the general case.

For the most part, the outstanding mathematical problems relevant
to the cold plasma model are boundary value problems for Maxwell's
equations. The dielectric tensor for these equations will render
them of elliptic type on one part of their domain and of
hyperbolic type on the remainder, except for a smooth curve (the
\emph{parabolic line}) separating the two regions. Little is known
about the formulation of well-posed boundary value problems for
equations which change type in this way, especially as the
equations that arise in the cold plasma model appear to have
certain fundamental differences from those that arise in gas
dynamics.

Careful reasoning about the mathematical properties of plasma
models is not needed merely in order to prevent ``mathematicians'
nightmares." An example is known \cite{otMSW} in which the boundary
conditions suggested by physical reasoning about the plasma lead
to a mathematically ill-posed problem in the expected function
space. Moreover, numerical experiments tend to confirm the
difficulties that arise when the model equations are subjected to
classical analytic techniques; see  \cite{otMSW} and
various remarks in  \cite{otW1}.

High-frequency waves can be modelled via geometrical optics. (Any
propagating electromagnetic field will tend to have high frequency
relative to the characteristic plasma frequencies; see,
\emph{e.g.}, Sec.\ 2.4 of  \cite{otMD}.) Mathematical
problems that arise in the geometrical optics approximation are
quite different from those that arise from applying Maxwell's
equations directly, and we do not pursue the geometrical optics
approach in this review. The complexity of the geometrical optics
approximation is due to significant difference in magnitude among
the terms of the plasma conductivity tensor at lower hybrid
frequencies; see  \cite{otRd} and the references therein.

The physics presented in Secs.\ \ref{otsec1.2.1} and
\ref{otsec1.2.2} essentially goes back to the work of Tonks and
Langmuir \cite{otTL} in the late 1920s. The results of Sec.\
\ref{otsec1.2.3} were already well known in the
1950s \cite{otAl,otAs,otSS}; those of Secs.\ \ref{otsec1.2.4} and
\ref{otsec1.2.5} date from the 1970s \cite{otLM,otPF} and 1980s,
respectively. In particular, Sec.\ \ref{otsec1.2.5} derives some
fundamental analytic formulas introduced by Weitzner in \cite{otW1} and \cite{otW2}; see also  \cite{otGW}.
Section \ref{otsec1.3} is based on recent results, \cite{otO1,otO2}
which extend analogous research on equations of Tricomi type ---
particularly  \cite{otLMP} and \cite{otLMPE}; see also
 \cite{otY}, an earlier paper which is based on \cite{otM1}.

In the sequel, a subscripted variable denotes (usually partial)
differentiation in the direction of the variable, whereas
subscripted numbers denote components of a matrix, vector, or
tensor. Differentiation of vector or matrix components in the
direction of a variable is indicated by preceding the subscripted
variable by a comma. Unless otherwise stated, a cartesian
coordinate system is assumed in which the subscript 1 denotes a
component projected onto the $x$-axis, the subscript 2 denotes a
component projected onto the $y$-axis, and the subscript 3 denotes
a component projected onto the $z$-axis. In particular,
$\mathbf{x} = \left(x_1,x_2,x_3\right)=\left(x,y,z\right)$ and we
denote the canonical cartesian basis by
$\left(\hat{\imath},\hat{\jmath},\hat{k}\right).$

\section{The cold plasma model}\label{otsec1.2}

A \emph{plasma} is a fluid composed of electrons and one or more
species of ions. Because it is a fluid, its evolution must satisfy
the equations of fluid dynamics. But because the particles of the
fluid are charged, they act as sources of an electromagnetic
field, which is governed by Maxwell's equations. The presence of
this intrinsic field leads to highly nonlinear behavior. Indeed,
the dominance of long-range electromagnetic interactions over the
short-range interatomic or intermolecular forces is often cited as
the defining characteristic of the plasma state.

If the plasma is at zero temperature, then Amontons' Law implies
that the pressure term in the equations for fluid motion will also
be zero, and the laws of fluid dynamics will enter only through
the conservation laws for mass and momentum. In fact, because
collisions can be neglected, the fluid aspect of the medium can be
virtually ignored. The plasma is then represented as a static
dielectric through which electromagnetic waves propagate.

In particular, \emph{zero-order quantities} --- the plasma
density, proportions of ions to electrons, and the background
magnetic field --- can all be considered static in time and
uniform in space. \emph{First-order quantities} --- the electric
field $\mathbf{E}$ and particle velocities $\mathbf{v},$ are
assumed to be expressible as plane waves: sinusoidal waves
proportional to functions having the form
$\exp\left[i\left(\mathbf{k}\cdot \mathbf{r}-\omega
t\right)\right],$ where $\mathbf{k}$ is the propagation vector of
the wave (not to be confused with the notation for the cartesian
basis vector $\hat{k}$); $\mathbf{r}$ is the radial coordinate in
space; $\omega$ is angular frequency; $i^2=-1.$ Thus in cartesian
coordinates, $\mathbf{k}\cdot \mathbf{r} = k_1x+k_2y+k_3z.$

In the following we review basic elements of the physical theory
that results from these assumptions. The material in Secs.\
\ref{otsec1.2.1}--\ref{otsec1.2.3} is completely standard and can
be found in many sources. The classical reference is Ch.\ 1 of
 \cite{otSt}; see also  \cite{otABB},
\cite{otBk}, \cite{otGi}, and Sec.\ 2 of \cite{otW2}. More recent surveys include nodes 43--45 of \cite{otFi} and Ch.\ 2 of  \cite{otSw}. A recent review
of theoretical plasma physics can be found in \cite{otBl}. We employ \emph{SI} units except where other units
are specified.

\subsection{Equations of motion}\label{otsec1.2.1}

Consider a single particle of mass $m,$ having charge $q=Z\delta
e,$ where $Z$ is a positive integer, $\delta$ equals 1 or $-1,$
and $e$ is the charge on an electron. Let the particle be
subjected only to the Lorentz force
\[
\mathbf{F}_L=q\left(\mathbf{E}+\mathbf{v}\times\mathbf{B}\right),
\]
where
\begin{equation}\label{otmag}
    \mathbf{B}=B_0\hat{k}.
\end{equation}
Equation (\ref{otmag}) implies that the applied magnetic field is
\emph{longitudinal}: its only nonzero component is directed along
the positive $z$-axis. (In fact, there is little harm in assuming,
somewhat more generally, that
\[
\mathbf{B} =B_0\hat{k} +\tilde{\mathbf
B}\left(x,y,z\right)\exp\left[i\left(\mathbf{k}\cdot
\mathbf{r}-\omega t\right)\right],
\]
with $|\tilde{\mathbf B}|<<|B_0|.$)

The equation of motion for the particle is given by Newton's
Second Law of Motion, that is,
\begin{equation}\label{otnewton}
    m\frac{d\mathbf{v}}{dt}=\mathbf{F}_L.
\end{equation}
In accordance with our assumption about first-order quantities, we
write
\[
\mathbf{v}\left(x,y,z,t\right)=\tilde{\mathbf{v}}\left(x,y,z\right)\exp\left[i\left(\mathbf{k}\cdot
\mathbf{r}-\omega t\right)\right],
\]
or
\[
\frac{d\mathbf{v}}{dt}=-i\omega\mathbf{v}.
\]
Substituting this result into (\ref{otnewton}) yields
\begin{equation}\label{otmot}
    -im\omega\tilde{\mathbf{v}}=q\left(\tilde{\mathbf{E}}+\tilde{\mathbf{v}}\times \mathbf{B}\right),
\end{equation}
where
\begin{equation}\label{otplanew}
    \mathbf{E}\left(x,y,z,t\right)=\tilde{\mathbf{E}}\left(x,y,z\right)\exp\left[i\left(\mathbf{k}\cdot
\mathbf{r}-\omega t\right)\right].
\end{equation}
Initially we will take $\tilde{\mathbf{E}}$ to be a constant
vector:
\begin{equation}\label{otEconst}
\tilde{\mathbf{E}}\left(x,y,z\right)=E_1\hat{\imath}+E_2\hat{\jmath}
+E_3\hat{k},
\end{equation}
where $E_1,$ $E_2,$ and $E_3$ are constants, and similarly for
$\tilde{\mathbf{v}}.$

Defining the \emph{cyclotron frequency}
\[
\Omega = \left|\frac{qB_0}{m}\right|,
\]
Eq.\ (\ref{otmot}) has solutions $\mathbf{v} =
\left(v_1,v_2,v_3\right)$ satisfying
\begin{eqnarray}
    v_1=\frac{iq}{m\left(\omega^2-\Omega^2\right)}\left(\omega
    E_1+i\delta\Omega E_2\right);\label{otv1}\\
    v_2=\frac{iq}{m\left(\omega^2-\Omega^2\right)}\left(\omega
    E_2-i\delta\Omega E_1\right);\label{otv2}\\
    v_3 = \frac{iq}{m\omega}E_3.\label{otv3}
\end{eqnarray}

\subsection{The dielectric tensor}\label{otsec1.2.2}

Although the above relations were derived for an individual
particle, they also hold, in our simplified linear model, for each
species of particle in a plasma consisting of electrons and $N-1$
species of ions. In particular, the plasma current can be written
as the sum
\begin{equation}\label{otcurr}
    \mathbf{j}=\sum_{\nu=1}^N n_\nu q_\nu\mathbf{v}_\nu,
\end{equation}
where $n_\nu$ is the density of particles having charge magnitude
$|q_\nu|=Z_\nu e.$

In the sequel we will only consider the aggregate of particles, in
which Eqs.\ (\ref{otmag})--(\ref{otv3}) pertain with the
quantities $\mathbf{v},$ $m,$ $q,$ $Z,$ $\delta,$ and $\Omega$
indexed by $\nu,$ where $\nu = 1, ..., N.$ Introduce the
\emph{electric displacement vector}
\begin{equation}\label{otdisplace1}
    \mathbf{D} = \mbox{vacuum displacement + plasma current} =
    \epsilon_0\mathbf{E}+\frac{i}{\omega} \mathbf{j},
\end{equation}
where $\epsilon_0$ is the permittivity of free space. It will be
convenient to express (\ref{otdisplace1}) in the form
\begin{equation}\label{otdisplace2}
    \mathbf{D} = \epsilon_0\mathbf{\mathbf{K}}\mathbf{E},
\end{equation}
where
\begin{equation}\label{otdisplace2a}
    D_i = \epsilon_0\sum_{j=1}^3K_{ij}E_j
\end{equation}
and $\mathbf{K} = \left(K_{ij}\right)$ is the \emph{dielectric
tensor} (also called the \emph{cold plasma conductivity tensor}).
The tensorial nature of this quantity reflects the anisotropy of
the plasma due to the presence of an applied magnetic field. (Note
that in the sequel the reader will be expected to distinguish
between the notation $\mathbf{K}$ for the dielectric tensor, the
notation $K_{ij}$ for the scalar element of its $i^{th}$ row and
$j^{th}$ column, and the notation $\mathcal{K}$ for the
type-change function of an elliptic-hyperbolic equation.)

Equations (\ref{otv1})--(\ref{otdisplace2}) imply that
\begin{equation}\label{otdiel}
    \mathbf{K} = \left(%
\begin{array}{ccc}
  s & -id & 0 \\
  id & s & 0 \\
  0 & 0 & p \\
\end{array}%
\right),
\end{equation}
where $s,$ $d,$ and $p$ are defined in terms of:

\emph{i)} the \emph{plasma frequency}, which for particles of the
$\nu^{th}$ species is given by
\[
\Pi_\nu^2 = \frac{n_\nu q^2}{\epsilon_0 m_\nu};
\]

\emph{ii}) the \emph{permittivities} $R$ or $L$ of a right- or
left-circularly polarized wave travelling in the direction
$\hat{k};$ these are given by
\[
R = 1 -
\sum_{\nu=1}^N\frac{\Pi_\nu^2}{\omega^2}\left(\frac{\omega}{\omega+\delta_\nu
\Omega_\nu}\right)
\]
and
\[
L =1 -
\sum_{\nu=1}^N\frac{\Pi_\nu^2}{\omega^2}\left(\frac{\omega}{\omega-\delta_\nu
\Omega_\nu}\right).
\]
In terms of these quantities,
\[
s = \frac{1}{2}\left(R+L\right),
\]
\[
d = \frac{1}{2}\left(R-L\right),
\]
and
\[
p =1- \sum_{\nu=1}^N\frac{\Pi_\nu^2}{\omega^2}.
\]
The mass of an electron is considerably smaller than the mass of
any ion; so the squared ion cyclotron frequencies obtained from
combining fractions in $R$ and $L$ can be neglected, leading to
the approximate formulas
\begin{equation}\label{otar}
    R \approx 1 -
\sum_{\nu=1}^{N-1}\frac{\Pi_e^2}{\omega^2+\omega\Omega_e+\Omega_e\Omega_{i_\nu}}
\end{equation}
and
\begin{equation}\label{otel}
    L\approx 1 -
\sum_{\nu=1}^{N-1}\frac{\Pi_e^2}{\omega^2-\omega\Omega_e+\Omega_e\Omega_{i_\nu}}.
\end{equation}
In these formulas, the subscripted $e$ denotes the value of the
relevant quantity for the electrons and the subscripted $i_\nu$
denotes that value for the $\nu^{th}$ species of ion. Note that,
by the same reasoning, the ion plasma frequencies can be neglected
in the definition of $p.$

\subsection{The plasma dispersion relation}\label{otsec1.2.3}

The field equations for the system described in Sec.\
\ref{otsec1.2.2} are Maxwell's equations,
\begin{equation}\label{otmaxw1}
    \nabla\times\mathbf{E} = -\frac{\partial \mathbf{B}}{\partial t},
\end{equation}
\begin{equation}\label{otmaxw2}
    \nabla\times\mathbf{B}=\mu_0\left(\mathbf{j}+\epsilon_0\frac{\partial\mathbf{E}}{\partial
    t}\right),
\end{equation}
where $\mu_0$ is the permeability of free space.

From the form of Eq.\ (\ref{otplanew}), it is clear that whenever
$\mathbf{E}$ and $\mathbf{B}$ are plane waves, Eqs.\
(\ref{otmaxw1}) and (\ref{otmaxw2}) reduce to the simpler form
\begin{equation}\label{otplanemaxw1}
    \mathbf{k}\times\mathbf{E} = \omega\mathbf{B}
\end{equation}
and
\begin{equation}\label{otplanemaxw2}
    \mathbf{k}\times\mathbf{B} =
    -i\mu_0\mathbf{j}-\omega\mu_0\epsilon_0\mathbf{E}.
\end{equation}
We can rewrite Eq.\ (\ref{otplanemaxw2}) to read
\begin{eqnarray}
\mathbf{k}\times\mathbf{B}
=-\omega\mu_0\left(\frac{i\mathbf{j}}{\omega}+\epsilon_0\mathbf{E}\right)
\nonumber\\
=-\omega\mu_0\mathbf{D}=-\epsilon_0\mu_0\omega\mathbf{K}\mathbf{E},
\label{otplanemaxw3}
\end{eqnarray}
where we have used (\ref{otdisplace1}) and (\ref{otdisplace2}) in
deriving the last identity.

Now using (\ref{otplanemaxw1}), (\ref{otplanemaxw3}), and the
elementary identity $\mu_0\epsilon_0=c^{-2},$ where $c$ is the
speed of light \emph{in vacuo}, we obtain
\[
\mathbf{k}\times\left(\mathbf{k}\times\mathbf{E}\right)
=\mathbf{k}\times\left(\omega\mathbf{B}\right)=\omega\left(\mathbf{k}\times\mathbf{B}\right)
\]
\[
=-\omega^2\epsilon_0\mu_0\mathbf{K}\mathbf{E} =
-\left(\frac{\omega}{c}\right)^2\mathbf{K}\mathbf{E},
\]
implying that
\begin{equation}\label{otdispers1}
    \mathbf{k}\times\left(\mathbf{k}\times\mathbf{E}\right)+\left(\frac{\omega}{c}\right)^2\mathbf{K}\mathbf{E}=0.
\end{equation}

Define the \emph{index of refraction vector}
\[
\mathbf{n} = \frac{c}{\omega}\mathbf{k}.
\]
With this construction, the scalar index of refraction acquires a
direction: that of the wave propagation vector $\mathbf{k}.$ In
terms of $\mathbf{n},$ Eq.\ (\ref{otdispers1}) reads
\begin{equation}\label{otdispers2}
    \mathbf{n}\times\left(\mathbf{n}\times\mathbf{E}\right)+\mathbf{K}\mathbf{E}=0.
\end{equation}
Conventionally, $\mathbf{k}$ (and thus $\mathbf{n}$) lies in the
$xz$-plane. Denote by $\theta$ the angle subtended by the vectors
$\mathbf{k}$ and $\mathbf{B}.$ Then (\ref{otdispers2}) can be
written as the matrix equation
\[
\left(%
\begin{array}{ccc}
  s-n^2\cos^2\theta & -id & n^2\cos\theta\sin\theta \\
  id & s-n^2 & 0 \\
  n^2\cos\theta\sin\theta & 0 & p - n^2\sin^2\theta \\
\end{array}%
\right)\left(%
\begin{array}{c}
  E_1 \\
  E_2 \\
  E_3 \\
\end{array}%
\right) = 0.
\]
This matrix equation has a nontrivial solution precisely when the
determinant of the $3\times3$ matrix vanishes. The condition for
the vanishing of that determinant, the \emph{cold plasma
dispersion relation} is, geometrically, the equation for the
wave-normal surface:
\begin{equation}\label{otwavenormal}
    An^4-Bn^2+C=0,
\end{equation}
where the coefficients satisfy
\begin{equation}\label{otA}
    A = s\sin^2\theta+p\cos^2\theta,
\end{equation}
\begin{equation}\label{otB}
    B = \left(s^2-d^2\right)\sin^2\theta+ps\left(1+\cos^2\theta\right),
\end{equation}
and
\begin{equation}\label{otC}
    C=p\left(s^2-d^2\right).
\end{equation}
Because the left-hand side of Eq.\ (\ref{otwavenormal}) is a
quadratic polynomial in $n^2$, we obtain from the quadratic
formula the solutions
\[
n^2=\frac{B\pm F}{2A}
\]
for $F$ satisfying $F^2=B^2-4AC.$ Using (\ref{otA})--(\ref{otC})
to write
\[
F^2 = \left(RL-ps\right)^2\sin^4\theta+4p^2d^2\cos^2\theta,
\]
we obtain
\[
    \tan^2\theta =
-\frac{p\left(n^2-R\right)\left(n^2-L\right)}{\left(sn^2-RL\right)\left(n^2-p\right)}.
\]
These equations yield criteria for \emph{cutoff}, where $n=0,$ or
\emph{resonance}, where $n\rightarrow\infty.$

Physically, cutoffs and resonances correspond to a change in the
behavior of the wave from possible propagation to evanescence.
Mathematically, we will identify certain resonances with a change
in type of the governing field equation, from hyperbolic (implying
wave propagation) to elliptic (implying evanescence). These
transitions may be accompanied, under certain conditions, by
reflection and/or absorption of the wave.

Sufficient conditions for cutoff are $p=0,$ $R=0,$ or $L=0$ ---
that is, a sufficient condition is $C=0.$ A sufficient condition
for resonance is $A=0$ which, given Eq.\ (\ref{otA}), can be
written
\begin{equation}\label{ottang}
    \tan^2\theta = -\frac{p}{s}.
\end{equation}
Particular cases of interest are $\theta = 0$ (propagation
parallel to the magnetic field) and $\theta = \pi/2$ (propagation
perpendicular to the magnetic field). We will be particularly
interested in the \emph{hybrid resonances} at $\theta = \pi/2,$
which occur at frequencies for which $s=0.$

\subsection{Electrostatic waves}\label{otsec1.2.4}

The electric field is said to be \emph{electrostatic} if it
approximately satisfies
\begin{equation}\label{otelectros}
    \mathbf{E}=-\nabla\Phi,
\end{equation}
where $\Phi$ is a scalar potential. Equation (\ref{otelectros}) is
satisfied locally by all time-independent electric fields and in
an ordinary dielectric, the converse is also true. However in cold
plasma there also exist time-dependent solutions of
(\ref{otelectros}). Cold plasma has been characterized as a linear dielectric
through which electromagnetic waves propagate. Thus these waves include, in
distinction to ordinary dielectrics, the special case of propagating
electrostatic waves.

We write $\Phi$ in the form
\[
\Phi(x,y,z;t)=\varphi\left(x,y,z\right)\exp\left[\mathbf{k}\cdot
\mathbf{r}-i\omega t\right],
\]
and add to Maxwell's equations (\ref{otmaxw1}), (\ref{otmaxw2})
the additional equation
\begin{equation}\label{otmax}
    div\,\mathbf{D} = 0,
\end{equation}
which follows from Gauss' law for electricity.

Equation (\ref{otelectros}) implies immediately that
\begin{equation}\label{otdsq}
    \nabla\times\mathbf{E}=0.
\end{equation}
This is most easily seen if we use differential forms, and note
that in terms of the exterior derivative, $\mathbf{E} = d\Phi,$ so
(\ref{otdsq}) is just the well-known property that
\[
d\mathbf{E}=d^2\Phi=0.
\]
(Here and below we will switch from vectors to forms whenever the
calculation is made more transparent thereby; but we will not
change notation for the underlying geometric object, making the
convention that the argument of the exterior derivative is always
assumed to be a differential form.) In either the vector or form
notation, identity (\ref{otdsq}) follows from the equality of
mixed partial derivatives. Applying the arguments relating
(\ref{otmaxw1}) to (\ref{otplanemaxw1}) and (\ref{otmaxw2}) to
(\ref{otplanemaxw2}) (\emph{translation into Fourier modes}), we
rewrite (\ref{otdsq}) in the form
\[
\mathbf{k}\times\mathbf{E} = 0.
\]
This implies, by the properties of the cross product, the
geometric fact that the vectors $\mathbf{k}$ and $\mathbf{E}$ are
parallel. We say that electrostatic waves are \emph{longitudinal}.
Physically, they appear as oscillations along the axis of the
magnetic field.

Thus we conclude that \emph{transverse} waves, which propagate in
a direction perpendicular to the magnetic field, must satisfy
\[
\mathbf{k}\cdot\mathbf{E} = 0.
\]
Again, differential forms are illuminating: The above identity
becomes $\delta d\Phi=0,$ where $\delta$ is the formal adjoint of
the exterior derivative $d.$ But $\Phi$ is a 0-form, so
$\delta\Phi=0$ automatically, and we find that transverse waves
satisfy
\[
\delta d\Phi+d\delta\Phi \equiv\Delta\Phi = 0,
\]
that is, transverse waves are necessarily harmonic.

\subsubsection{Plane-layered media}\label{otsec1.2.4.1}

If we allow a plane-layered inhomogeneous medium (parameterized by
$x$), the electrostatic potential has the form
\[
\Phi\left(x,y,z\right) = \varphi(x)\exp\left[i\left( k_2 y+  k_3
z\right)\right],
\]
where $k_j$ is the $j^{th}$ component of the wave vector
$\mathbf{k}$ for $j=1,2,3.$ Substitution of this form for the
electric potential into Eq.\ (\ref{otmax}) yields, using
(\ref{otdisplace2a}), the single scalar equation \cite{otLM}
\begin{equation}\label{otode}
    K_{11}\varphi_{xx}+\left(K_{11,x}+i\sigma_0\right)\varphi_x=0,
\end{equation}
where
\[
\sigma_0= k_3\left(K_{13}+K_{31}\right)+
k_2\left(K_{12}+K_{21}\right),
\]
and zero-order terms in $\varphi$ have been neglected. This
equation has a power-series solution except where $K_{11}$
vanishes.

Explicit solutions of the model equation (\ref{otode}) under
various physical assumptions are given in Sec.\ 1 of \cite{otPF}, the Appendix to  \cite{otLM}, and Sec.\ C
of  \cite{otGW}.

It is easy to believe that inhomogeneities may develop in a
plasma. For example, if the temperature is not exactly zero, the
difference in velocity between electrons and ions can be expected
to destabilize an initially homogeneous distribution. However, it
is difficult to imagine a force that will restrict these
inhomogeneities to a 1-parameter foliation, which would be
necessary in order to arrive at Eq.\ (\ref{otode}). Formally, an
electromagnetic potential leading to Eq.\ (\ref{otode}) could be
induced by applying a driving potential to the metallic plates of
a condenser. But in practice, this plasma geometry has little
application either in the laboratory or in nature.

\subsubsection{A two-dimensional inhomogeneity}\label{otsec1.2.4.2}

Suppose instead that the medium is a cold, anisotropic plasma with
a two-dimensional inhomogeneity parameterized by two variables,
$x$ and $z.$ Then the field potential has the form
\[
\Phi\left(x,y,z\right) =
\varphi\left(x,z\right)\exp\left[ik_2y\right].
\]
The electric field $\mathbf{E}$ is then given by
\[
E = -\nabla\Phi = \left(E_1,E_2,E_3\right) =
-\left(\varphi_xe^{ik_2y},i k_2\varphi
e^{ik_2y},\varphi_ze^{ik_2y}\right).
\]

Maxwell's equations for the electric displacement vector
$\mathbf{D}=\left(D_1,D_2,D_3\right)$ take the form
\begin{equation}\label{otmaxwell}
    0 = \nabla\cdot\mathbf{D}=D_{1,x}+D_{2,y}+D_{3,z}.
\end{equation}
We continue to neglect those terms which do not contain
derivatives of $\varphi,$ as $\varphi$ is assumed to oscillate
rapidly.

Because neither $\varphi$ nor $K_{ij}$ have any dependence on $y,$
the problem is effectively two-dimensional. Applying Eq.\
(\ref{otdisplace2a}), the surviving terms of Eq.\
(\ref{otmaxwell}) are (setting $\epsilon_0$ equal to 1)
\[
    D_{1,x}=-\left[K_{11}\varphi_{xx}+K_{11,x}\varphi_x+K_{12}\varphi_xi
k_2+K_{13}\varphi_{zx}+K_{13,x}\varphi_z\right]e^{ik_2y};
\]
\[
D_{2,y}=-\left[K_{21}\varphi_xi k_2+K_{23}\varphi_zi k_2
\right]e^{ik_2y};
\]
\[
D_{3,z}=-\left[K_{31}\varphi_{xz}+K_{31,z}\varphi_x+K_{32}i
k_2\varphi_z+K_{33}\varphi_{zz}+K_{33,z}\varphi_z\right]e^{ik_2y}.
\]
Collecting terms, we find that \cite{otPF}
\begin{equation}\label{otpde}
    K_{11}\varphi_{xx}+2\sigma\varphi_{xz}+K_{33}\varphi_{zz}+\alpha_1\varphi_x+\alpha_2\varphi_z=0,
\end{equation}
where
\[
2\sigma = K_{13}+K_{31};
\]
\[
\alpha_1 = K_{11,x}+i k_2\left(K_{12}+K_{21}\right)+K_{31,z};
\]
\[
\alpha_2=K_{13,x}+i k_2\left(K_{23}+K_{32}\right)+K_{33,z}.
\]

Two-dimensional inhomogeneities of the kind represented by Eq.\
(\ref{otpde}) can be expected to arise in toroidal fields, such as
those created in tokamaks.

The entries of the matrix $K$ under our assumptions on
$\mathbf{B}_0$ imply that $\sigma = 0,$ so we can write Eq.\
(\ref{otpde}) in the form
\begin{equation}\label{otcp}
    K_{11}\varphi_{xx}+K_{33}\varphi_{zz}+ \mbox{lower-order
    terms}= 0.
\end{equation}
Equation (\ref{otcp}) is of either elliptic or hyperbolic type,
depending on whether the sign of the product
\begin{equation}\label{otprod}
    K_{11}\cdot K_{33}=\left(1-\sum_{\nu=1}^N\frac{\Pi_\nu^2}{\omega^2-\Omega_\nu^2}\right)\cdot\left(1 -
    \sum_{\nu=1}^N\frac{\Pi_\nu^2}{\omega^2}\right)
\end{equation}
is, respectively, positive or negative.

The sign of $K_{11}$ changes at the \emph{cyclotron} resonances
$\omega^2=\Omega_\nu^2.$ The cold plasma model breaks down at
these resonances, as three terms of the dielectric tensor become
infinite. The sign of $K_{11}$ also changes at the \emph{hybrid}
resonances, at which
\begin{equation}\label{otsonic1}
    1=\sum_{\nu=1}^N\frac{\Pi_\nu^2}{\omega^2-\Omega_\nu^2}.
\end{equation}
(These resonances, which have both a low-frequency and a
high-frequency solution, are hybrid in that they involve both
plasma and cyclotron frequencies.) In particular, the sign changes
at the \emph{lower} hybrid resonance,
\begin{equation}\label{otsonic2}
    1+\frac{\Pi_e^2}{\Omega_e^2}=\frac{\Pi_i^2}{\omega^2},
\end{equation}
where as before, the subscript $e$ denotes electron frequency, and
the subscript $i$ denotes ion frequency. At the hybrid resonance
frequencies the cold plasma model retains its validity.

The sign of $K_{33}$ changes on the surface
\begin{equation}\label{otsonic3}
    1 = \sum_{\nu=1}^N\frac{\Pi_\nu^2}{\omega^2},
\end{equation}
the resonance at which the frequency of the applied wave equals
the plasma frequency of the medium. We may suppose that an
electromagnetic wave propagating through a plasma does so at a
much higher frequency than any of the characteristic frequencies
of the plasma. Otherwise, the plasma magnetic field would prevent
the waves from propagating very far (\emph{c.f.} \cite{otKu}). Thus in evaluating (\ref{otprod}) and in the
sequel we will take $K_{33}$ to be strictly positive.

Borrowing the terminology of fluid dynamics, we will refer to
resonances such as (\ref{otsonic1})--(\ref{otsonic3}) as
\emph{sonic conditions} on Eq.\ (\ref{otcp}).

\subsubsection{The type of the governing equation}\label{otsec1.2.4.3}

In order to understand the possible variants of Eq.\ (\ref{otcp}),
we consider the coordinate transformation
$\left(x,z\right)\rightarrow\left(\xi\left(x,z\right),\eta\left(x,z\right)\right),$
where
\[
\xi = K_{11}\left(x,z\right).
\]
In these coordinates, the higher-order terms of Eq.\ (\ref{otcp})
assume the form
\begin{eqnarray}
K_{11}\varphi_{xx}+K_{33}\varphi_{zz}=\left(\xi\xi_x^2+K_{33}\xi_z^2\right)\varphi_{\xi\xi}+\nonumber\\
\left(\xi\xi_x\eta_x+K_{33}\xi_z\eta_z\right)\varphi_{\xi\eta}+\left(\xi\eta_x^2+K_{33}\eta_z^2\right)\varphi_{\eta\eta}.\label{ottrans}
\end{eqnarray}
In order that the transformation
$\left(x,z\right)\rightarrow\left(\xi,\eta\right)$ be nonsingular,
we require that its Jacobian be nonvanishing, \emph{i.e.,}
\begin{equation}\label{otjac}
    \xi_x\eta_z-\xi_z\eta_x\ne 0.
\end{equation}
Because we want the coefficients of the cross term
$\varphi_{\xi\eta}$ to be zero in the new coordinates, we impose
the condition that
\begin{equation}\label{otcross}
    \xi\xi_x\eta_x+K_{33}\xi_z\eta_z=0.
\end{equation}
Both $\xi$ and $K_{33}$ are given, and it is easy for the two
first derivatives of $\eta$ to satisfy (\ref{otjac}) and
(\ref{otcross}) simultaneously.

Two possibilities arise. Either

\medskip

\emph{i)} $\xi$ and $\xi_z$ never vanish simultaneously, or

\medskip

\emph{ii)} there exist one or more points $\left(x,z\right)$ on
the domain at which
\begin{equation}\label{otsi}
    \xi\left(x,z\right)=\xi_z\left(x,z\right)=0.
\end{equation}

In case \emph{i)}, the condition $\xi=0$ implies, via
(\ref{otcross}) and the assumption that $K_{33}$ is positive, the
accompanying condition $\eta_z=0.$ But if $\xi$ and $\eta_z$ both
vanish, then the coefficient of $\varphi_{\eta\eta}$ in
(\ref{ottrans}) vanishes; that is,
\[
\xi\eta_x^2+K_{33}\eta_z^2 = 0
\]
whenever $\xi=0.$ Again using (\ref{otcross}), we obtain from
Eqs.\ (\ref{otcp}) and (\ref{ottrans}) an equation with
higher-order terms having the form
\begin{equation}\label{otTT1}
    \varphi_{\xi\xi}+\frac{\xi\eta_x^2+K_{33}\eta_z^2}{\xi\xi_x^2+K_{33}\xi_z^2}\varphi_{\eta\eta}=0.
\end{equation}
The denominator in the coefficient of $\varphi_{\eta\eta}$ cannot
be zero: $\xi$ and $\xi_z$ cannot vanish simultaneously, and if
$\xi_x$ vanishes, then $\xi_z$ must be nonzero in order to
preserve condition (\ref{otjac}). So Eq.\ (\ref{otTT1}) is of the
form
\begin{equation}\label{otTT2}
    \varphi_{\xi\xi}+\mathcal{K}\left(\xi,\eta\right)\varphi_{\eta\eta}=0,
\end{equation}
where $\mathcal{K}\left(\xi,\eta\right)=0$ if and only if $\xi=0,$
an equation of \emph{Tricomi type}.

In case \emph{ii)}, condition (\ref{otjac}) prevents $\eta_z$ from
vanishing when $\xi_z$ vanishes. Thus if $\xi$ and $\xi_z$ vanish
together, the coefficient of $\varphi_{\eta\eta}$ in
(\ref{ottrans}) will not vanish at that point. Thus in case
\emph{ii)} we obtain from (\ref{otcp}), (\ref{ottrans}), and
(\ref{otcross}) an equation with higher-order terms having the
form
\begin{equation}\label{otKT1}
    \frac{\xi\xi_x^2+K_{33}\xi_z^2}{\xi\eta_x^2+K_{33}\eta_z^2}\varphi_{\xi\xi}+\varphi_{\eta\eta}=0,
\end{equation}
where the numerator in the coefficient of $\varphi_{\xi\xi},$ but
not the denominator, is zero whenever $\xi$ is zero. That is, Eq.\
(\ref{otKT1}) is an equation of the form
\begin{equation}\label{otKT2}
    \mathcal{K}\left(\xi,\eta\right)\varphi_{\xi\xi}+\varphi_{\eta\eta}
    = 0,
\end{equation}
where $\mathcal{K}\left(\xi,\eta\right)=0$ if and only if $\xi=0,$
an equation of \emph{Keldysh type}.

See Sec.\ 1.2 of  \cite{otBi},  \cite{otCC1},
Sec.\ 1 of  \cite{otMSW}, and Eqs.\ (75)--(78) of \cite{otW1} for arguments of this kind.

The regularity of weak solutions to equations of Tricomi type can
be established by microlocal arguments; see \cite{otGra}
and \cite{otGro} and, especially, \cite{otPa1} and
\cite{otPa2}. These arguments appear to fail for equations of
Keldysh type, and one expects weaker regularity for weak solutions
to such equations.

The crucial question is: does condition (\ref{otsi}) occur in our
physical model? The answer to that question is ``yes."

\subsubsection{Geometry of the resonance curve (\emph{after Piliya and Fedorov})}\label{otsec1.2.4.4}

Returning to our original $xz$-coordinates, we set the elements
$K_{11}$ and $K_{22}$ of the dielectric tensor equal to
$\mathcal{K},$ and the element $K_{33}$ equal to $\eta.$ The
coefficients of the only other nonzero elements, $K_{12}$ and
$K_{21},$ are zero in Eq.\ (\ref{otcp}), so only $K_{11}$ and
$K_{33}$ play a direct role in the analysis. The sonic condition
is equivalent to the alternative:
\begin{equation}\label{otson}
    \mathcal{K}=0
\end{equation}
or
\begin{equation}\label{otangle}
    \mathcal{K}\sin^2\theta + \eta\cos^2\theta = 0,
\end{equation}
where $\theta\left(x,z\right)$ is the angle between the direction
of $\mathbf{B_0}$ and the $xz$-plane; \emph{c.f.} (\ref{ottang}).

The singular points on the \emph{sonic line} (\ref{otson}) are the
points at which this curve (which is not a generally a line in
standard coordinates) is tangent to the projection of the force
lines of $\mathbf{B_0}$ in the $xz$-plane --- that is, the flux
lines of the magnetic field. The singular points of the graph
$\Gamma$ of Eq.\ (\ref{otangle}) are the points at which the flux
lines of $\mathbf{B_0}$ are normal to $\Gamma.$

This motivates the placement of the origin at a singular point of
the sonic line, with the $z$-axis (the axis along which
$\mathbf{B_0}$ is directed) tangent to the sonic line. The
$x$-axis is directed along the inward normal to the sonic line,
relative to the hyperbolic region of Eq.\ (\ref{otpde}). Then
$K_{11}$ and $\sigma$ both vanish at the origin. Taking both $x$
and $z$ to be small, one can write
\begin{equation}\label{otscale1}
    K_{11}=\frac{x}{a}+\frac{z^2}{b}
\end{equation}
and
\begin{equation}\label{otscale2}
    K_{33}=-\eta_0,
\end{equation}
where $\eta_0$ is a positive constant. Scale $x$ and $z,$ via
\begin{equation}\label{otdimvar1}
    x\rightarrow \tilde x = x/a
\end{equation}
and
\begin{equation}\label{otdimvar2}
    z \rightarrow \tilde z = z/a\sqrt{\eta_0},
\end{equation}
in order to obtain dimensionless variables $\tilde x$ and $\tilde
z.$ In this way, one obtains in place of (\ref{otpde}) the
equation
\begin{equation}\label{otPF}
    -\left(\tilde x+A\tilde z^2\right)\varphi_{\tilde x \tilde x} +\varphi_{\tilde z\tilde z}-\varphi_{\tilde x}=0,
\end{equation}
where $A$ is a constant, for the simple case in which the
coefficients of cross terms vanish identically \cite{otPF}.

\subsection{Analytic difficulties in the electromagnetic case (\emph{after H. Weitzner)}}\label{otsec1.2.5}

In this section we suppose that the electric field satisfies Eqs.\
(\ref{otplanew}) and (\ref{otEconst}), but no longer assume that
the electric field satisfies condition (\ref{otelectros}). Closely
following  \cite{otW1}, we attempt to study the resulting
field equations using conventional analytic tools, in order to see
what difficulties arise.

Repeating the calculations of Eqs.\
(\ref{otmaxw1})--(\ref{otdispers1}) in greater detail, we compute
\begin{eqnarray}
    \nabla\times\left(\nabla\times\mathbf{E}\right)=\nabla\times\left(-\frac{\partial
B}{\partial t}\right) =
\nabla\times\left(i\omega\mathbf{B}\right)=\nonumber\\
i\omega\left(\nabla\times\mathbf{B}\right)=i\omega\left[\mu_0\left(\mathbf{j}+\epsilon_0\frac{\partial\mathbf{E}}{\partial
t}\right)\right]=\nonumber\\
i\omega\mu_0\mathbf{j}+i\omega\mu_0\epsilon_0\left(-i\omega\mathbf{E}\right)=i\omega\mu_0\mathbf{j}+\omega^2\mu_0\epsilon_0\mathbf{E}\nonumber\\
=\mu_0\omega^2\left(\frac{i}{\omega}\mathbf{j}+\epsilon_0\mathbf{E}\right)=\mu_0\omega^2\mathbf{D}=\mu_0\epsilon_0\omega^2\mathbf{K}\mathbf{E}.\label{otgeneq}
\end{eqnarray}
Now
\[
\nabla\times\mathbf{E} = \left(E_{3,y} -
E_{2,z}\right)\hat{\imath} +
\left(E_{1,z}-E_{3,x}\right)\hat{\jmath} +
\left(E_{2,x}-E_{1,y}\right)\hat{k}.
\]
It is obvious that this quantity vanishes identically in the
electrostatic case: apply (\ref{otelectros}) and equate mixed
partial derivatives. But if $\nabla\times\mathbf{E}$ is itself a
gradient, then the quantity
$\nabla\times\left(\nabla\times\mathbf{E}\right)$ vanishes for the
general case as well. We will understand the seriousness of this
latter difficulty once we evaluate the left-hand side of Eq.\
(\ref{otgeneq}). Explicitly,
\[
\nabla\times\left(\nabla\times\mathbf{E}\right)=\left(E_{2,xy}-E_{1,yy}-E_{1,zz}+E_{3,xz}\right)\hat{\imath}
+
\]
\[
\left(E_{3,yz}-E_{2,zz}-E_{2,xx}+E_{1,yx}\right)\hat{\jmath} +
\left(E_{1,zx}-E_{3,xx}-E_{3,yy}+E_{2,zy}\right)\hat{k}.
\]
Applying (\ref{otplanew}), (\ref{otEconst}) to the right-hand
side, we obtain the algebraic expression
\[
    \left[k_1k_2E_2-\left(k_2^2+k_3^2\right)E_1+k_1k_3E_3\right]\hat{\imath}
    + \left[k_2k_3E_3-\left(k_3^2+k_1^2\right)E_2+k_{21}E_1\right]\hat{\jmath}
\]
\begin{equation}\label{otFmode}
+\left[k_{31}E_1-\left(k_1^2+k_2^2\right)E_3+k_{32}E_2\right]\hat{k}.
\end{equation}
This object can be written as the matrix operator
\[
L\mathbf{E} = \left(%
\begin{array}{ccc}
  -\left(k_2^2+k_3^2\right) & k_1k_2 & k_1k_3 \\
  k_2k_1 & -\left(k_3^2+k_1^2\right) & k_2k_3 \\
  k_3k_1 & k_3k_2 & -\left(k_1^2+k_2^2\right) \\
\end{array}%
\right)
\left(%
\begin{array}{c}
  E_1 \\
  E_2 \\
  E_3 \\
\end{array}%
\right).
\]
The system (\ref{otgeneq}) is uniquely solvable if and only if the
operator $L$ can be inverted --- that is, if and only if
\[
\det\left(%
\begin{array}{ccc}
  -\left(k_2^2+k_3^2\right) & k_1k_2 & k_1k_3 \\
  k_2k_1 & -\left(k_3^2+k_1^2\right) & k_2k_3 \\
  k_3k_1 & k_3k_2 & -\left(k_1^2+k_2^2\right) \\
\end{array}%
\right)\ne 0.
\]
But it is easy to check that this determinant vanishes identically
for all $\left(k_1,k_2,k_3\right).$ Of course (\ref{otFmode}) is
just a translation of
$\nabla\times\left(\nabla\times\mathbf{E}\right)$ into Fourier
mode. Because the symbol of a differential operator is a natural
generalization of the idea of Fourier modes, we can interpret the
foregoing computation to mean that the symbol of the differential
operator $\nabla\times\left(\nabla\times\,\,\right)$ vanishes
identically. This is a serious obstacle to understanding
(\ref{otgeneq}). As Weitzner notes in  \cite{otW2},
neither the type of Eq.\ (\ref{otgeneq}) (which is given by the
sign of the symbol) nor the order of the equation (which is given
by the degree of the symbol) are determined by standard analytic
methods.

\subsubsection{Choices of potential and gauge}\label{otsec1.2.5.1}

It is therefore necessary to impose an additional hypothesis. A
natural one is that an electromagnetic potential exists. But in
distinction to the electrostatic case, we do not assume that
$\mathbf{E}$ can be derived by simply taking the negative gradient
of a scalar field.

In order to compare our computations with the extensive
expositions in  \cite{otW1} and \cite{otW2} we adopt,
only in Secs.\ \ref{otsec1.2.5.1} and \ref{otsec1.2.5.2}, the
convention that the time-harmonic dependence is of the form
$\exp\left[i\omega t\right]$ in units of $c/\omega.$ (This is in
distinction to (\ref{otplanew}).) Because our time derivatives
usually end up being taken twice, this only has an effect on the
sign in a few intermediate calculations. However, with this sign
convention, Maxwell's equations for plane waves assume the
slightly different form
\begin{equation}\label{otmaxvar1}
    \nabla\times\mathbf{E} = -i\mathbf{B},
\end{equation}
\begin{equation}\label{otmaxvar2}
    \nabla\times\mathbf{B} = i\mathbf{D} = i\mathbf{K}\mathbf{E}.
\end{equation}

The first choice of potentials is to let the vector $\mathbf{A}$
denote the magnetic potential and to introduce a second, scalar
potential, $\Phi.$ We then write
\begin{equation}\label{otmagpoten}
    \mathbf{B} = \nabla\times\mathbf{A}
\end{equation}
and
\begin{equation}\label{otpot1}
    \mathbf{E} = -i\mathbf{A}-\nabla\Phi.
\end{equation}
Taking the curl of the second equation, we obtain
\begin{equation}\label{otinterm}
    \nabla\times\mathbf{E}=-i\nabla\times\mathbf{A}-\nabla\times\nabla\Phi.
\end{equation}
Evaluating the last term on the right-hand side of
(\ref{otinterm}) using differential forms, $\Phi$ is a zero-form
and
\begin{equation}\label{otdouble}
    \nabla\times\nabla\Phi=d^2\Phi=0.
\end{equation}
Equations (\ref{otmagpoten}) and (\ref{otdouble}) imply that
(\ref{otmaxvar1}) is satisfied under condition (\ref{otpot1}) for
any smooth choice of $\mathbf{A}$ and $\Phi.$

Notice that we automatically obtain from hypothesis
(\ref{otmagpoten}) an extra condition
\[
\nabla\cdot\mathbf{B}=\nabla\cdot\left(\nabla\times\mathbf{A}\right),
\]
which is to say, in terms of differential forms, that the 2-form
$\mathbf{B}$ and the 1-form $\mathbf{A}$ satisfy
\[
d\mathbf{B}=d^2\mathbf{A}=0.
\]

In order to evaluate (\ref{otmaxvar2}), we notice, continuing to
interpret the magnetic potential $\mathbf{A}$ as a 1-form and
$\Phi$ as a zero-form, that if $g$ is defined by
\[
g\left(\mathbf{A}\right)=\mathbf{A}+df,
\]
where $f$ is a smooth 0-form, then
\[
d\left(g\left(\mathbf{A}\right)\right)=d\left(\mathbf{A}+df\right)=d\mathbf{A}+d^2f=d\mathbf{A}=\mathbf{B},
\]
so the magnetic field remains invariant under the \emph{gauge
transformation} $g.$ Moreover, because $\delta f = 0$ for any
zero-form $f,$ we have
\[
\Delta f = -\left(\delta d + d \delta\right)f = -\delta d f.
\]
Thus, given any smooth potential $\mathbf{A},$ we can choose $f$
to satisfy the Poisson equation
\[
\Delta f = \delta \mathbf{A},
\]
in which case
\[
\delta\left(g\left(\mathbf{A}\right)\right)=\delta\left(\mathbf{A}+df\right)=\delta\mathbf{A}-\Delta
f = 0.
\]
We call the gauge produced by such a $g$ a \emph{Coulomb}
(\emph{transverse}, \emph{radiation}, or \emph{Hodge}) gauge. In
vector notation,
\[
i\nabla\cdot g\left(\mathbf{A}\right) = 0.
\]
Computing (\ref{otmaxvar2}) in the Coulomb gauge, we
obtain \cite{otW1}
\[
\Delta g\left(\mathbf{A}\right)-i\mathbf{K}\nabla\Phi +
\mathbf{K}g\left(\mathbf{A}\right)=0.
\]
Computing the symbol $\sigma$ of this operator by the same method
that was applied to the operator
$\nabla\times\left(\nabla\times\mathbf{E}\right),$ we find that
$\sigma=-|\mathbf{k}|^4\left(\mathbf{K}\mathbf{k}\right)\cdot\mathbf{k},$
a polynomial of degree six in $\mathbf{k}.$ That the corresponding
system is of order six is an expected result for a system of two
first-order equations for vectors in $\mathbb{R}^3.$

Replacing the Coulomb gauge by a slightly more complicated gauge
in which
\[
i\nabla\cdot\left(\mathbf{K}^\ast\mathbf{A}\right) =0,
\]
where the superscripted asterisk denotes the adjoint matrix, we
obtain a self-adjoint operator in (\ref{otmaxvar2}). This more
complicated gauge can be constructed by the same general method
that led to the Coulomb gauge, provided that we solve a slightly
more complicated Poisson problem. The symbol corresponding to this
self-adjoint operator can also be calculated by the methods
introduced earlier, and that symbol is also a sixth-degree
polynomial in $\mathbf{k}.$

However, we can obtain a fourth-order system, which is more
convenient for analysis, if we impose an additional hypothesis:
that the plasma has axisymmetric geometry. While this is a very
strong hypothesis, it is satisfied by plasmas produced in
tokamaks.

In order to motivate the choice of potential in this case, we make a
few preliminary calculations. Only for the remainder of this
section, the subscripts $r,$ $\theta,$ and $z$ when affixed to a
vector are to be interpreted as the radial, angular, and axial
components of the vector unless preceded by a comma; if preceded by
a comma, they are to be considered partial derivatives in the
direction of the component. (The subscripted-variable notation for
partial derivatives of scalar functions remains unchanged.) Adopting
the basis
\[
\mathbf{u}_r = \cos\theta\hat{\imath}+\sin\theta\hat{\jmath},
\]
\[
\mathbf{u}_\theta =
-\sin\theta\hat{\imath}+\cos\theta\hat{\jmath},
\]
\[
\mathbf{u}_z = -\hat{k},
\]
we recall that in the axisymmetric case we can write
\[
\mathbf{E} =
\left(E_r\left(r,z\right)\mathbf{u}_r+E_\theta\left(r,z\right)r\mathbf{u}_\theta+E_z\left(r,z\right)\mathbf{u}_z\right)e^{im\theta},
\]
and similarly for $\mathbf{B}.$ If $m=0,$ the waves preserve the
axisymmetry of the underlying static plasma medium, as the wave
vector satisfies $ \mathbf{k} = \left(k_r,0,k_z\right).$ We will
restrict our attention to this simple special case, in which
\begin{eqnarray}
    \nabla\times\mathbf{E} =
\frac{1}{r}\left[E_{z,\theta}\mathbf{u}_r+E_{r,z}\left(r\mathbf{u}_\theta\right)+\left(rE_\theta\right)_r\mathbf{u}_z\right] \nonumber\\
-\frac{1}{r}\left[E_{z,r}\left(r\mathbf{u}_\theta\right)+rE_{\theta,z}\mathbf{u}_r+E_{r,\theta}\mathbf{u}_z\right]=\nonumber\\
-E_{\theta,z}\mathbf{u}_r+\left(E_{r,z}-E_{z,r}\right)\mathbf{u}_\theta+\frac{1}{r}\left(rE_\theta\right)_{,r}\mathbf{u}_z.\label{otcylbas}
\end{eqnarray}
Thus (\ref{otmaxvar1}) implies in particular that
\begin{equation}\label{otmagbas1}
    -iE_{\theta,z} = B_r
\end{equation}
and
\begin{equation}\label{otmagbas2}
    \frac{\left(rE_\theta\right)_{,r}}{r} = -iB_z.
\end{equation}
Just as Eqs.\ (\ref{otmagbas1}) and (\ref{otmagbas2}) imply, using
(\ref{otmaxvar1}), that $B_r$ and $B_z$ can each be expressed in
terms of derivatives of $E_\theta,$ so it is possible to use
(\ref{otmaxvar2}) to show that the other cylindrical components of
$\mathbf{E}$ and $\mathbf{B}$ can be expressed as appropriate
derivatives of $E_\theta$ and $B_\theta.$ This will allow the
angular components of $\mathbf{E}$ and $\mathbf{B}$ to play the
role of potentials for the two fields.

Applying (\ref{otmaxvar1}) to the middle term of the last identity
in (\ref{otcylbas}) yields
\begin{equation}\label{otcylfield1}
    E_{r,z}-E_{z,r}=-iB_\theta.
\end{equation}
Because $\mathbf{E}$ and $\mathbf{B}$ have exactly analogous forms
and the left-hand and middle terms of (\ref{otmaxvar2}) is exactly
analogous to (\ref{otmaxvar1}) with a change of sign, we can
immediately write
\[
D_r=iB_{\theta,r}
\]
and
\[
D_z=-i\frac{\left(rB_\theta\right)_{,z}}{r}.
\]
Now the extreme right-hand side of (\ref{otmaxvar2}) yields $E_r$
and $E_z$ (see Eqs.\ (22), (23) of  \cite{otW1}) and one
obtains
\begin{equation}\label{otcylfield2}
    B_{r,z}-B_{z,r}=iD_\theta=i\left(K_{\theta
    r}E_r+K_{\theta\theta}E_\theta+K_{\theta z}E_z\right),
\end{equation}
completing the system of equations for $E_\theta$ and $B_\theta.$

 \subsubsection{Variational interpretation}\label{otsec1.2.5.2}

 Continuing to adopt the special hypotheses and special notation of
 Sec.\ \ref{otsec1.2.5.1}, we continue to review the analysis in  \cite{otW1} of
 geometry-preserving plane waves in an axisymmetric plasma.

 Equations (\ref{otcylfield1}), (\ref{otcylfield2}) can be associated
 to an energy functional:
 \begin{eqnarray}
    \mathcal{E} =
    \int\{\left[\nabla\left(rE_\theta^\ast\right)\cdot\nabla\left(rE_\theta\right)\right]/r^2+\left[\nabla\left(rB_\theta^\ast\right)\cdot\mathbf{K}\nabla\left(rB_\theta\right)\right]/r^2\Delta\nonumber\\
    +iE_\theta\left[\left(rB_\theta^\ast\right)_{,r}\left(K_{zr}K_{r\theta}-K_{z\theta}K_{rr}\right)/r+B_{\theta,z}^\ast\left(K_{r\theta}K_{zz}-K_{rz}K_{z\theta}\right)\right]/\Delta\nonumber\\
    -iE_\theta^\ast\left[\left(rB_\theta\right)_{,r}\left(K_{zr}K_{\theta r}-K_{\theta z}K_{rr}\right)/r+B_{\theta,z}\left(K_{\theta r}K_{zz}-K_{zr}K_{\theta
    z}\right)\right]/\Delta\nonumber\\
    -B_\theta^\ast B_\theta-E_\theta^\ast E_\theta\left[\det\left(\mathbf
    K\right)\right]/\Delta\}r\,dr dz, \label{otenfun}
\end{eqnarray}
where
\[
\nabla = \frac{\partial}{\partial r}{r}+\frac{\partial}{\partial
z}{z},
\]
and
\[
\Delta = K_{rr}K_{zz}-K_{rz}K_{zr}.
\]
Provided $\mathbf{K}$ can be made self-adjoint, so can
$\mathcal{E}.$ Form a right-handed orthogonal set
$\left(\mathbf{v}, \mathbf{\theta},\mathbf{u}\right),$ where
\[
\mathbf{u} = \sin\beta r + \cos\beta  k
\]
and
\[
\mathbf{v} = \cos\beta r - \sin\beta  k.
\]
The basis is to be chosen so that $\mathbf{u}$ lies in the
poloidal direction and $\mathbf{v}$ lies orthogonal to it; so we
write the magnetic field vector in the form
\[
\mathbf{B} = B_0\left[\cos\alpha\theta+\sin\alpha\left(\sin\beta r
+\cos\beta z\right)\right],
\]
where $\alpha,$ $\beta,$ and $B_0$ depend only on $r$ and $z.$ In
this notation, the variational equations of $\mathcal{E}$ form a
second-order system in which the differential operator for one of
the equations is essentially the Laplacian $\mathcal{L}.$ We
ignore that equation, as standard analytic methods can be applied
to it. The differential operator for the other equation looks like
\begin{equation}\label{otlaplus}
    \mathcal{L}+\left(\mathbf{u}\cdot\nabla\right)^2,
\end{equation}
and that is the equation --- in particular, the second of the two
differential operators in that equation --- that we will study in
the remainder of this review. The term (\ref{otlaplus}) in Eq.\
(\ref{otenfun}) can be written explicitly, in terms of the chosen
basis, in the form
\begin{eqnarray}
r\nabla\cdot\left[\left(\frac{\xi}{r^2\Delta}\right)\nabla\left(rB_\theta\right)\right]-r\nabla\cdot\left[\left(\frac{\zeta\sin^2\alpha}{r^2\Delta}\right)\left(\mathbf{{u}}\cdot\nabla\right)\left(rB_\theta\right)\mathbf{{u}}\right]\nonumber\\
-i\theta\cdot\nabla\left(rB_\theta\right)\times\nabla\left(\frac{\mu\cos\alpha}{r\Delta}\right)+B_\theta=\nonumber\\
\left(r\Delta\right)^{-1}\left[\mu\left(\zeta-\xi\right)\sin\alpha\mathbf{{u}}\cdot\nabla\left(rE_\theta\right)+
i\left(\mu^2-\xi\zeta\right)\sin\alpha\cos\alpha\mathbf{{v}}\cdot\nabla\left(rE_\theta\right)\right],\label{otwz72}
\end{eqnarray}
where
\[
\xi = 1+\sum_{\nu=1}^N\frac{\Pi_\nu^2}{\Omega_\nu^2-\omega^2},
\]
\[
\zeta = \xi + \sum_{\nu=1}^N\frac{\Pi_\nu^2}{\omega^2}-1,
\]
and
\[
\mu = \sum_{\nu =
1}^N\frac{\Pi_\nu^2\Omega_\nu}{\omega\left(\Omega_\nu^2-\omega^2\right)}.
\]

Equation (\ref{otwz72}) is only elliptic for negative values of
$\xi.$ Physically, this is the condition for so-called
\emph{lower-hybrid} frequencies, at which
\[
1+\frac{\Pi_e^2}{\Omega_e^2}<\frac{\Pi_i^2}{\omega^2},
\]
\emph{c.f.} (\ref{otsonic2}). Noticing that $\xi$ is a function of
$r$ and $z,$ define a new variable $\eta\left(r,z\right)$ so the
curves $\eta = $ constant are orthogonal to the curves $\xi=$
constant. Rewriting (\ref{otwz72}) in
$\left(\xi,\eta\right)$-coordinates, the behavior of the solution
depends on whether or not
\[
\mathbf{u}\cdot\nabla\xi =0.
\]
This identity implies that flux surfaces coincide with resonance
surfaces. In that case, Eq.\ (\ref{otwz72}) is analogous to Eq.\
(\ref{otKT2}) of Sec.\ \ref{otsec1.2.4.3} and is not of Tricomi
type. The second-order terms of that equation can be written in
the form
\begin{equation}\label{otxioper}
    L(u) = f\left(\xi,\eta\right)\left[\xi
u_{\xi\xi}+M\left(\xi,\eta\right)u_{\eta\eta}\right],
\end{equation}
where $u=u\left(\xi,\eta\right)$ is a scalar function; $f$ and $M$
are given well behaved scalar functions near $\xi=0$ and, in
addition, $M$ is positive.

The physical model allows two further alternatives: If the curve
representing the flux surface in two dimensions is collinear with
the resonance curve as in (\ref{otxioper}), then the plasma can be
treated as a perpendicularly stratified medium, which is
essentially the case considered in Sec.\ \ref{otsec1.2.4.1}. If
the resonance curve is tangent to the curve representing the flux
surface, then we are in a more mathematically and physically
interesting case. In this latter case, the simplest model for the
operator $L$ of (\ref{otwz72}) is an operator for which the
highest-order terms have the form
\begin{equation}\label{otcpop}
    \tilde{L}(u) = \left(x-y^2\right)u_{xx}+u_{yy}.
\end{equation}
Note that this operator is closely related to the differential
operator of Eq.\ (\ref{otPF}). The two operators can be made
virtually identical by replacing the coordinate transformation
(\ref{otdimvar1}) by
\begin{equation}\label{otdimvar1a}
    x\rightarrow \tilde x = -x/a.
\end{equation}

\subsection{A conjecture about the singular set}\label{otsec1.2.6}

Methods for deriving the smoothness of solutions to the Tricomi
equation appear to fail for an operator of the form (\ref{otcpop})
whenever the domain includes the origin of coordinates. This
suggests the existence of a singular point at the origin, a
conjecture which is supported by an analysis of characteristic
lines.

In order for a characteristic line to pass through the origin, the
point$\left(x,y\right)$ would need to satisfy the identity
\begin{equation}\label{otorig}
x = \lambda y^2
\end{equation}
for some constant $\lambda,$ and also the characteristic equation
for (\ref{otcpop}). Substituting (\ref{otorig}) into the
characteristic equation
\begin{equation}\label{otchrt}
    \left(x-y^2\right)dy^2+dx^2=0,
\end{equation}
one obtains the equation
\begin{equation}\label{otcharor}
    \frac{dy^2}{\left(2\lambda
ydy\right)^2}=\frac{1}{\left(1-\lambda\right)y^2},
\end{equation}
or
\[
4\lambda^2+\lambda-1=0.
\]
This polynomial has two real solutions; considering that the
characteristic equation (\ref{otcharor}) has two roots, one
concludes \cite{otMSW} that four characteristic lines must pass
through the origin --- two more than pass through any other
hyperbolic point. This motivates the suspicion that solutions of
at least the equation $\tilde{L}u=0$ will tend to be singular at
the origin. It has been observed that an energy sink or plasma
heating zone might be associated with such a singularity; see
 \cite{otGW}, \cite{otMSW}, \cite{otPF},
\cite{otW1}, and \cite{otW2} for details on this and other
issues raised in this section.

\section{Analysis of the model equation}\label{otsec1.3}

Physical reasoning suggests that the \emph{closed} Dirichlet
problem, in which data are prescribed along the entire boundary of
the domain, should be well-posed for the cold plasma model.
However, the closed Dirichlet problem has been shown to be
ill-posed, in the classical sense, for the equation
\[
\left(x-y^2\right)u_{xx}+u_{yy}+\frac{1}{2}u_x=0
\]
on a typical domain \cite{otMSW}. This leads us to ask whether a
well-posed problem with closed boundary data can be formulated in
a suitably weak sense. In this section we address the ``existence"
part of that question.

Because the operator introduced in Eq.\ (\ref{otcpop}) is not of
Tricomi type at the origin, where it satisfies a condition of the
form (\ref{otsi}), we expect weaker regularity than we have for
operators which are uniformly of Tricomi type. In particular,
although we can show the existence of very weak solutions in
$L^2,$ we do not expect $H^1$ regularity for the closed Dirichlet
problem. This lack of optimism is supported by numerical
experiments \cite{otMSW}. Moreover, our methods are insufficient to
determine the uniqueness of a solution.

Denote by $\Omega$ a bounded, connected domain of $\mathbb{R}^2$
having piecewise smooth boundary $\partial\Omega,$ oriented in a
counterclockwise direction; the domain includes both an arc of the
sonic curve and the origin of coordinates in $\mathbb{R}^2.$ (This
insures that our equation will be elliptic-hyperbolic but not
equivalent to an equation of Tricomi type.)

Define \cite{otLMP}, for a given $C^1$ function
$\mathcal{K}\left(x,y\right),$ the space
$L^2\left(\Omega;|\mathcal{K}|\right)$ and its dual. These spaces
consist, respectively, of functions $u$ for which the norm
\[
||u||_{L^2\left(\Omega;|\mathcal{K}|\right)}=\left(\int\int_\Omega|\mathcal{K}|u^2dxdy\right)^{1/2}
\]
is finite, and functions $u\in L^2\left(\Omega\right)$ for which the
norm
\[
||u||_{L^2\left(\Omega;|\mathcal{K}|^{-1}\right)}=\left(\int\int_\Omega|\mathcal{K}|^{-1}u^2dxdy\right)^{1/2}
\]
is finite. Standard arguments allow us to define the space
$H^1_0(\Omega; \mathcal{K})$ as the closure of $C_0^\infty(\Omega)$
with respect to the norm
\begin{equation}\label{otH101}
    ||u||_{H^1(\Omega; \mathcal{K})}=\left[\int\int_{\Omega}
\left(|\mathcal{K}|u_x^2+u_y^2+u^2\right)\,dxdy\right]^{1/2}.
\end{equation}
The $H^1_0(\Omega; \mathcal{K})$-norm has the explicit form
\begin{equation}\label{otH102}
    ||u||_{H^1_0(\Omega; \mathcal{K})}=\left[\int\int_\Omega
\left(|\mathcal{K}|u_x^2+u_y^2\right)\,dxdy\right]^{1/2},
\end{equation}
which can be derived from (\ref{otH101}) via a weighted Poincar\'e
inequality.

In the following we denote by $C$ generic positive constants, the
value of which may change from line to line.

\subsection{The closed Dirichlet problem for distribution
solutions}\label{otsec1.3.1}

Consider the equation
\begin{equation}\label{otf-alt1}
    Lu=f,
\end{equation}
where $f$ is a given, sufficiently smooth function of $(x,y)$ and
\begin{equation}\label{otf-alt2}
    L=\left(x-y^2\right)\frac {\partial^2} {\partial x^2}+\frac{\partial^2}{\partial y^2}+ \kappa\frac{\partial}{\partial x}
\end{equation}
for a given constant $\kappa.$ By a \emph{distribution solution}
of equations (\ref{otf-alt1}), (\ref{otf-alt2}) with the boundary
condition
\begin{equation}\label{otboundary}
    u(x,y)=0\,\forall (x,y)\in\partial\Omega
\end{equation}
we mean a function $u\in L^2(\Omega)$ such that $\forall \xi \in
H^1_0(\Omega;\mathcal{K})$ for which $L^\ast\xi\in L^2(\Omega),$ we
have
\begin{equation}\label{otds}
    \left(u,L^\ast\xi\right)=\langle f,\xi \rangle.
\end{equation}
Here $L^\ast$ is the adjoint operator; $(\,,\,)$ denotes the $L^2$
inner product on $\Omega;$ $\langle \,,\, \rangle$ is the
\emph{duality bracket} associated to the $H^{-1}$ norm
\[
||w||_{H^{-1}(\Omega;\mathcal{K})}=\sup_{0\neq\xi\in
C^\infty_0(\Omega)}\frac{|\langle w,\xi
\rangle|}{||\xi||_{H^1_0(\Omega;\mathcal{K})}}.
\]
Such solutions have also been called \emph{weak;} \emph{c.f.} Eq.\
(2.13) of  \cite{otBe}, Sec.\ II.2. In fact they are a
little smoother than generic distribution solutions, as they lie
in a classical function space.

The existence of solutions to boundary value problems can be shown
to follow from \emph{energy inequalities} having the general form
\[
||v||_V\leq C||L^\ast v||_U,
\]
where $U$ and $V$ are suitable function spaces. We will combine
such an inequality with the Riesz Representation Theorem to prove
the existence of distribution solutions; see  \cite{otBe}
for a general reference.

\bigskip

\noindent\textbf{Lemma 3.1. ( \cite{otO2}).} \emph{The
inequality}
\[
||u||_{H^1_0(\Omega;\mathcal{K})}\leq C||Lu||_{L^2(\Omega)},
\]
\emph{is satisfied for $u\in C^2_0(\Omega),$ where the positive
constant $C$ depends on $\Omega$ and $\mathcal{K}$; $L$ is defined
by (\ref{otf-alt2}) with $\kappa \in \left[0,2\right];$
$\mathcal{K}=x-y^2.$}

\medskip

\begin{proof} We outline the proof; for details, see  \cite{otO2}, Sec.\ 2. Initially, let $1\leq\kappa\leq 2,$
and let $\delta$ be a small, positive constant. Define an operator
$M$ by the identity
\[
    Mu=au+bu_x+cu_y
\]
for $a=-1,$ $c=2\left(2\delta-1\right)y,$ and
\[
b= \left\{
        \begin{array}{cr}
    \exp\left(2\delta \mathcal{K}/Q_1\right) & \mbox{if $\left(x,y\right)\in\Omega^+$} \\

    \exp\left(6\delta \mathcal{K}/Q_2\right) & \mbox{if $\left(x,y\right)\in\Omega^-$}\\
    \end{array}
    \right.,
\]
where
$\Omega^+=\left\lbrace\left(x,y\right)\in\Omega\,|\,\mathcal{K}
> 0\right\rbrace$ and $\Omega^-=\Omega\backslash\Omega^+.$ Choose
$Q_1=\exp\left(2\delta\mu_1\right),$ where
$\mu_1=\max_{\left(x,y\right)\in\overline{\Omega^+}}\mathcal{K}.$
Define the negative number
$\mu_2=\min_{\left(x,y\right)\in\overline{\Omega^-}}\mathcal{K}$
and let $Q_2=\exp\left(\mu_2\right).$ Notice that $b\leq Q_1$ on
$\Omega^+$ and $b> Q_2$ on $\Omega^-.$

We will estimate the quantity $\left(Mu,Lu\right)$ from above and
below. As in the Tricomi case \cite{otLMP}, one of the coefficients
in $Mu$ fails to be continuously differentiable on all of
$\Omega.$ When integrating this quantity, a cut should be
introduced along the line $\mathcal{K}=0.$ The boundary integrals
involving $a,$ $b,$ and $c$ on either side of this line will
cancel.

The boundary terms vanish by the compact support of $u.$
Integration by parts yields the identity
\[
\left(Mu,Lu\right)=\int\int_{\Omega^+\cup\Omega^-}\alpha
u_x^2+2\beta u_xu_y+\gamma u_y^2\,dxdy,
\]
where
\[
\alpha =
\left(\frac{c_y}{2}-a-\frac{b_x}{2}\right)\mathcal{K}+\left(\kappa-\frac{1}{2}\right)b-cy,
\]
for
\[
\alpha_{|\Omega^+}=\left(2 -\frac{b}{Q_1}\right)\delta
\mathcal{K}+2\left(1-2\delta\right)y^2+\left(\kappa-\frac{1}{2}\right)b
\]
and
\[
\alpha_{|\Omega^-} =
\left(3\frac{b}{Q_2}-2\right)\delta|\mathcal{K}|+2\left(1-2\delta\right)y^2+\left(\kappa-\frac{1}{2}\right)b;
\]
\[
\beta=\frac{1}{2}\left[c\left(\kappa-1\right)-b_y\right] = \left\{
        \begin{array}{cr}
    y\left[2\delta \left(b/Q_1\right)+\left(\kappa-1\right)\left(2\delta-1\right)\right]\leq |y| & \mbox{in $\Omega^+$} \\

    y\left[6\delta \left(b/Q_2\right)+\left(\kappa-1\right)\left(2\delta-1\right)\right]\leq \kappa|y| & \mbox{in $\Omega^-$}\\
    \end{array}
    \right.;
\]
\[
\gamma=\frac{1}{2}\left(b_x-c_y\right)-a =\left\{
\begin{array}{cr}
    2\left(1-\delta\right)+\delta\left(b/Q_1\right) & \mbox{in $\Omega^+$} \\

    2\left(1-\delta\right)+3\delta\left(b/Q_2\right) & \mbox{in $\Omega^-$}\\
    \end{array}
    \right..
\]
On $\Omega^+,$ for any scalars $\xi$ and $\eta,$ we have
\[
\alpha\xi^2+2\beta\xi\eta+\gamma\eta^2\geq
\alpha\xi^2-\left(y^2\xi^2+\eta^2\right)+\gamma\eta^2=
\]
\[
\left[\left(2-\frac{b}{Q_1}\right)\delta
\mathcal{K}+\left(1-4\delta\right)y^2+\left(\kappa-\frac{1}{2}\right)b\right]\xi^2+\left[\left(1-2\delta\right)+\frac{6b}{Q_1}\right]\eta^2
\]
\[
\geq \delta\left(\mathcal{K}\xi^2+\eta^2\right),
\]
provided $\delta$ is sufficiently small. On $\Omega^-,$
\[
\alpha\xi^2+2\beta\xi\eta+\gamma\eta^2\geq
\alpha^2\xi^2-2\left(y^2\xi^2+\eta^2\right)+\gamma\eta^2=
\]
\[
\left[\left(3\frac{b}{Q_2}-2\right)\delta|\mathcal{K}|-4\delta
y^2+\left(\kappa -
\frac{1}{2}\right)b\right]\xi^2+\delta\left(3\frac{b}{Q_2}-2\right)\eta^2
\]
\[
\geq \delta\left(|\mathcal{K}|\xi^2+\eta^2\right).
\]
Arguing in this way on each subdomain (and taking $\xi=u_x,$ $\eta
= u_y$), we obtain
\begin{equation}\label{otestlow}
    \left(Mu,Lu\right)\geq\delta||u||_{H^1_0\left(\Omega;\mathcal{K}\right)}^2.
\end{equation}
For the upper estimate, we have \cite{otLMP}
\begin{equation}\label{otestup}
    \left(Mu,Lu\right)\leq ||Mu||_{L^2}||Lu||_{L^2}\leq
    C\left(K,\Omega\right)||u||_{H^1_0(\Omega;\mathcal{K})}||Lu||_{L^2(\Omega)}.
\end{equation}
Combining (\ref{otestlow}) and (\ref{otestup}), and dividing
through by the weighted $H_0^1$-norm of $u,$ completes the proof
for the case $\kappa\in\left[1,2\right].$

Now let $0\leq\kappa <1.$ Again subdivide the domain into $\Omega^+$
and $\Omega^-$ by introducing a cut along the curve $K=0.$ Integrate
by parts, choosing $a=-1;$
\[
b=\left\{
\begin{array}{cr}
    -N\mathcal{K} & \mbox{in $\Omega^+$} \\

    N\mathcal{K} & \mbox{in $\Omega^-$}\\
    \end{array}
    \right.,
\]
where $N$ is a constant satisfying
\begin{equation}\label{otrange}
    \frac{1+\tilde\delta}{3-\kappa}<N<\frac{1-\tilde\delta}{\kappa+1}
\end{equation}
for a sufficiently small positive constant $\tilde\delta;$
$c=-4Ny.$ The boundary integrals involving $a$ and $c$ on either
side of the cut will cancel and the boundary integrals involving
$b$ will be zero along the cut. Inequality (\ref{otestlow}) can be
derived by an argument broadly analogous to the case
$\kappa\in\left[1,2\right].$ Applying (\ref{otestup}) then
completes the proof. \end{proof}

\begin{theorem} [ \cite{otO2}] \label{ottheorem1} The Dirichlet problem
(\ref{otf-alt1}), (\ref{otf-alt2}), (\ref{otboundary}) with
$\kappa\in[0,2]$ possesses a distribution solution $u\in
L^2(\Omega)$ for every $f \in H^{-1}(\Omega;\mathcal{K}).$
\end{theorem}

\begin{proof} Again, we only outline the proof (\emph{c.f.}  \cite{otLMP}, Theorem 2.2). Define for
$\xi \in C_0^\infty$ a linear functional
\[
J_f(L\xi)=\langle f, \xi \rangle.
\]
This functional is bounded on a subspace of $L^2$ by the inequality
\begin{equation}\label{otschwartz}
    \left |\langle f, \xi \rangle\right| \leq
    ||f||_{H^{-1}\left(\Omega;\mathcal{K}\right)}||\xi||_{H_0^1\left(\Omega;\mathcal{K}\right)}
\end{equation}
and by applying Lemma 3.1 to the second term on the right.
Precisely, $J_f$ is a bounded linear functional on the subspace of
$L^2\left(\Omega\right)$ consisting of elements having the form
$L\xi$ with $\xi\in C_0^\infty\left(\Omega\right).$ Extending
$J_f$ to the closure of this subspace by Hahn-Banach arguments, we
obtain a functional defined on all of $L^2.$ The Riesz
Representation Theorem then guarantees the existence of a
distribution solution in the self-adjoint case. If $\kappa\ne1,$
then $L$ is not self-adjoint, and
\begin{equation}\label{otf-adj}
    L^\ast=\left(x-y^2\right)\frac {\partial^2} {\partial x^2}+\frac{\partial^2}{\partial y^2}+ \left(2-\kappa\right)\frac{\partial}{\partial
    x}.
\end{equation}
Estimating $L$ for $\kappa$ in $[0,2]$ will also yield estimates
for the adjoint $L^\ast.$ Applying the preceding argument to the
adjoint operator completes the proof of Theorem \ref{ottheorem1}.
\end{proof}

\subsection{Mixed boundary value problems with closed boundary
data}\label{otsec1.3.2}

It is also possible to form \emph{mixed Dirichlet-Neumann}
problems for operators of the form (\ref{otf-alt2}). Mixed
boundary value problems arise in various contexts in plasma
physics (\emph{e.g.},  \cite{otBF}) and in related topics
from electromagnetic theory (\emph{e.g.},  \cite{otLF},
which is related to the model of Sec.\ \ref{otsec1.2.4.1}).
However, the results of this section also imply --- by taking the
set of boundary points on which the Dirichlet conditions are
imposed to be empty --- the existence of weak solutions to a class
of Neumann problems.

Denote by $\mathbf{u}=\left(u_1,u_2\right)$ and
$\mathbf{w}=\left(w_1,w_2\right)$ measurable vector-valued
functions on $\Omega.$ Define $\mathcal{H_\mathfrak{K}}$ to be the
Hilbert space of measurable functions on $\Omega$ for which the
norm induced in the obvious way by the weighted $L^2$ inner
product
\[
\left(\mathbf{u},\mathbf{w}\right)_\mathfrak{K}=\int\int_\Omega\left(|\mathcal{K}|u_1w_1+u_2w_2\right)dxdy
\]
is finite. In the notation for these spaces, $\mathfrak{K}$
denotes a diagonal matrix having entries $|\mathcal{K}|$ and 1.

By a \emph{weak solution} of a mixed boundary-value problem in
this context we mean an element $\mathbf{u} \in
\mathcal{H}_\mathfrak{K}(\Omega)$ such that
\begin{equation}\label{otweak1}
    -\left(\mathbf{u},\mathcal{L}^\ast\mathbf{w}\right)_{L^2\left(\Omega;\mathbb{R}^2\right)}=\left(\mathbf{f},\mathbf{w}\right)_{L^2\left(\Omega;\mathbb{R}^2\right)}
\end{equation}
for every function $\mathbf{w}\in
C^1\left(\overline{\Omega};\mathbb{R}^2\right)$ for which
$\mathfrak{K}^{-1}\mathcal{L}^\ast\mathbf{w}\in
L^2\left(\Omega;\mathbb{R}^2\right)$ and for which
\begin{equation}\label{otweak2}
    w_1=0\,\forall \left(x,y\right)\in G
\end{equation}
and
\begin{equation}\label{otweak3}
    w_2=0\,\forall \left(x,y\right)\in \partial\Omega\backslash G,
\end{equation}
where $G\subset\partial\Omega.$ Choose the differential operator
$\mathcal{L}$ to have the form
\begin{equation}\label{otop}
\left(
\begin{array}{cc}
  \mathcal{K}\partial_x & \partial_y \\
  \partial_y & -\partial_x \\
\end{array}%
\right)+\left(%
\begin{array}{cc}
  \kappa & 0 \\
  0 & 0 \\
\end{array}%
\right),
\end{equation}
where $\kappa$ is a number in $\left[0,1\right].$

\bigskip

\begin{theorem} [ \cite{otO2}] \label{ottheorem2} Let $G$ be a subset of
$\partial\Omega$ and let $\mathcal{K}=x-y^2.$ Define the functions
$b\left(x,y\right)=m\mathcal{K}+s$ and $c(y)=\mu y-t,$ where $\mu$
is a positive constant,
\[
m= \left\{
        \begin{array}{cr}
    \left(\mu + \delta\right)/2 & \mbox{in $\Omega^+$} \\

    \left(\mu - \delta\right)/2 & \mbox{in $\Omega^-$}\\
    \end{array}
    \right.
\]
for a small positive number $\delta,$ and $t$ is a positive
constant such that $\mu y - t < 0 \, \forall y \in \Omega.$ Let
$s$ be a sufficiently large positive constant. In particular,
choose $s$ to be so large that the quantities $m\mathcal{K}+s,$
$2cy+s,$ and $b^2+\mathcal{K}c^2$ are all positive. Let
\begin{equation}\label{otstar1}
    bdy-cdx\leq 0
\end{equation}
\emph{on $G$ and}
\begin{equation}\label{otstar2}
    \mathcal{K}\left(bdy-cdx\right) \geq 0
\end{equation}
on $\partial\Omega\backslash G.$ Then there exists for every
$\mathbf{f}$ such that $\mathfrak{K}^{-1}\mathcal{M}^T
\mathbf{f}\in L^2(\Omega)$ a weak solution to the mixed
boundary-value problem (\ref{otweak1})--(\ref{otweak3}) for
$\mathcal{L}$ given by Eq.\ (\ref{otop}) with $\kappa=0,$ where
the superscripted $T$ denotes matrix transpose.
\end{theorem}

\begin{proof} We give the idea of the proof \cite{otO2}. One shows that there exists a positive
constant $C$ such that
\[
\left(\mathbf{\Psi},\mathcal{L^\ast M}\mathbf{\Psi}\right)\geq
C\int\int_\Omega\left(|\mathcal{K}|\Psi_1^2+\Psi_2^2\right)dxdy
\]
for any sufficiently smooth 2-vector $\mathbf{\Psi},$ provided
conditions (\ref{otweak2}), (\ref{otweak3}) are satisfied on the
boundary for $\mathbf{w}=\mathcal{M}\mathbf\Psi,$ where
$\mathcal{L^\ast}$ is given by (\ref{otop}) with $\kappa=1$ and
\[
\mathcal{M} = \left(%
\begin{array}{cc}
  b & c \\
  -\mathcal{K}c & b \\
\end{array}%
\right).
\]
This inequality leads to an application of the Riesz
Representation Theorem by arguments which are roughly analogous to
those used to prove Theorem \ref{ottheorem1}.
\end{proof}

\bigskip

Despite the technical nature of the hypotheses in Theorem
\ref{ottheorem2}, simple domains which satisfy them are very easy
to construct --- \emph{e.g.}, a box in the first quadrant having a
vertex at the origin of coordinates, or a narrow lens about the
sonic curve in the first quadrant. Note that by taking $G$ to be
the empty set, we obtain a solution to the closed conormal problem
(\emph{c.f.}  \cite{otPi}). But in order for Theorem
\ref{ottheorem2} to guarantee a solution to the closed Dirichlet
problem, we would need to find a domain on which $G$ could be
taken to be the entire boundary; it is not obvious how to
construct such a domain. And, as was also the case in Theorem
\ref{ottheorem1}, the proof of Theorem \ref{ottheorem2} does not
establish the uniqueness of solutions.

In addition to its intrinsic mathematical and physical interest,
the formulation of boundary value problems illuminates other
topics in the analysis of the cold plasma model. For example, it
is shown in Sec.\ \ref{otsec1.2.4.3}, by a tedious analytic
argument, that away from the origin the governing equation for the
model is of Tricomi type, whereas in the neighborhood of the
origin it is of Keldysh type. This distinction is also suggested,
without reference to such terminology, by other analytic arguments
in  \cite{otPF} and in Sec.\ 4 of  \cite{otW2}. If
we try to form a standard elliptic-hyperbolic boundary value
problem in which the hyperbolic region is composed of intersecting
characteristics, we might choose both these characteristics to
originate at points on the arc of the resonance curve $x=y^2$ that
lies in the first quadrant, or both of them to lie in the fourth
quadrant. We then obtain a standard problem for a
vertical-ice-cream-cone-shaped region (in the former case, the
ice-cream cone is held upside down), similar to those formulated
for the Tricomi equation (Eq.\ (\ref{otTT2}) with
$\mathcal{K}\left(\xi,\eta\right)=\xi$). The domain geometry is
exactly analogous to, for example, Fig.\ 2 of \cite{otM2}, with the line $AB$ in that figure replaced by an
arc of the curve $x=y^2,$ lying either completely above or
completely below the $x$-axis. But the origin will not be
included, as that is a singular point of the characteristic
equation (\ref{otchrt}). If we include the origin, we are led to a
hyperbolic region bounded by characteristics in the second and
third quadrants, a horizontal-ice-cream-cone-shaped region similar
to those formulated for the Cinquini-Cibrario equation (Eq.\
(\ref{otKT2}) with $\mathcal{K}\left(\xi,\eta\right)=\xi$). In
this case typical domain geometry is analogous to Fig.\ 2 of
\cite{otCC2}, with the line $MN$ in that figure replaced by an
arc of the curve $x=y^2$ which is symmetric about the $x$-axis;
see also Remark \emph{i)} following Corollary 11 of
\cite{otO2}. Thus the defining analytic character of the
equation is clearly apparent in the geometry of the natural
boundary value problems.

\end{document}